\documentclass{PoS}

\def\beq{\begin{equation}}
\def\eeq{\end{equation}}
\def\bea{\begin{eqnarray}}
\def\eea{\end{eqnarray}}
\def\beqn{\begin{eqnarray}} \def\eeqn{\end{eqnarray}}
\def\beeq{\begin{eqnarray}}
\def\eeeq{\end{eqnarray}}

\def\ep{\epsilon}

\def\nn{\nonumber}

\def\td#1{\tilde{\delta}\left(#1\right)}

\title{
\vspace*{-2.0cm}
Four-dimensional regularization of higher-order computations: FDU approach}

\ShortTitle{4D regularization of H.O. computations: FDU approach}

\author{\speaker{Germ\'an F. R. Sborlini}$^{\ a,b}$, F\'elix Driencourt-Mangin$^b$, Roger J. Hern\'andez-Pinto$^c$ and Germ\'an Rodrigo$^b$\\\\
        $^a$Dipartimento di Fisica, Universit\`a di Milano and INFN Sezione di Milano,
I-20133 Milan, Italy.\\\
        $^b$Instituto de F\'{\i}sica Corpuscular, Universitat de Val\`{e}ncia -- 
Consejo Superior de Investigaciones Cient\'{\i}ficas, Parc Cient\'{\i}fic, E-46980 Paterna, Valencia, Spain.\\\
        $^c$Facultad de Ciencias F\'isico-Matem\'aticas, Universidad Aut\'onoma de Sinaloa, Ciudad Universitaria, CP 80000, Culiac\'an, Sinaloa, M\'exico.\\\

        E-mail: \email{german.sborlini@unimi.it}}

\abstract{We have recently proposed a new regularization framework based on the loop-tree duality theorem. This theorem allows to rewrite loop level amplitudes in terms of tree-level structures and phase-space integrations. In consequence, it is possible to combine naturally real and virtual contributions at integrand level. Moreover, through the introduction of a proper momentum-mapping, a complete local cancellation of infrared singularities is achieved, by-passing the necessity of counter-terms. In this article, we briefly explain the implementation of this novel approach to compute some physical processes, and we show how to deal with both infrared and ultraviolet divergences without using DREG.}

\FullConference{EPS-HEP 2017, European Physical Society conference on High Energy Physics\\
		5-12 July 2017\\
		Venice, Italy}

\begin{document}

\section{Introduction}
\label{sec:introduction}
The presence of ill-defined expressions in intermediate steps of QFT computations requires the introduction of regularization methods to render them convergent and unambiguously defined. In many cases, these problems arise as a consequence of physical singularities, such as infrared (IR) or ultraviolet (UV) ones. Within the community of high-energy physics, one of the most accepted methods is Dimensional Regularization (DREG). Roughly speaking, DREG assumes that the number of space-time dimensions is extended from $d=4$ to $d=4-2\ep$; thus, the convergence issues manifest as $\epsilon$-poles.

On the other hand, we know that any physically relevant observable must be independent of the regularization technique applied. In particular, this means that infrared-safe observables in QCD (and calculated within DREG) have to be finite in the limit $\ep \to 0$. However, these kind of computations require to consider both virtual (i.e. loop diagrams) and real (i.e. diagrams with additional physical particles being radiated) contributions. Provided that the loop contributions have been properly renormalized, then only IR singularities will survive in the virtual component: the Kinoshita-Lee-Nauenberg (KLN) theorem \cite{Kinoshita:1962ur,Lee:1964is} establishes that these IR singularities can be canceled by adding those present in the real contributions\footnote{Strictly speaking, the presence of additional singularities -such as initial state collinear configurations- requires to include additional subtraction counter-terms to cancel them. These counter-terms are related with the factorization theorem and the perturbative evolution of PDFs.}. Thus, the combination of real and virtual components must be free of IR singularities, which translates into the absence of $\epsilon$-poles after integrating them within the DREG approach. This is the main reason behind the development of the subtraction methods \cite{Kunszt:1992tn,Frixione:1995ms,Catani:1996jh,Catani:1996vz}, which introduce proper counter-terms in the real and virtual terms in order to cancel separately the IR-singularities: i.e. these counter-terms exactly reproduce the IR-singular behaviour of the real matrix elements, and their integrated form must cancel the IR poles present in the virtual components. There are several variations of the subtraction framework, that provide alternative paths to build the counter-terms. In any case, the cancellation of singularities in the virtual component takes place \emph{after} integration, even if the counter-terms render the real contribution locally integrable. 

In this work, we explain how to directly use the real-emission amplitude as a counter-term for the renormalized virtual contribution: this constitutes the central idea of the four-dimensional unsubtraction (FDU) approach \cite{Hernandez-Pinto:2015ysa,Sborlini:2015uia,Sborlini:2016fcj,Sborlini:2016gbr,Rodrigo:2016hqc,Sborlini:2016hat,Hernandez-Pinto:2016uwx}. Within this framework, the introduction of IR counter-terms is avoided and a fully local cancellation of singularities is achieved. In consequence, the limit $\ep \to 0$ can be safely considered at integrand level and a complete four-dimensional numerical implementation becomes feasible.

\section{Description of the FDU framework}
\label{sec:FDUframework}
The four-dimensional unsubtraction (FDU) approach is a fully-local, four-dimensional regularization framework to implement higher-order computations in any quantum field theory. It relies on the loop-tree duality (LTD) theorem \cite{Catani:2008xa,Rodrigo:2008fp,Bierenbaum:2010cy}, which establishes the possibility of decomposing any loop amplitude into tree-level objects by \emph{cutting} (i.e. \emph{putting on-shell}) internal lines circulating the loops. As an example, let's consider a generic one-loop scalar Feynman integral with $N$ external legs, whose momenta are denoted $\{p_i\}_{i=1\ldots N}$; the application of the LTD decomposition leads to
\beq
\int_{\ell} \, \prod_{i=1}^N \frac{1}{q_i^2-m_i^2+\imath 0} = - \sum_{i=1}^N \, \int_\ell \td{q_i} \, \prod_{j\neq i, j=1}^N \frac{1}{q_j^2-m_j^2-\imath 0 \, \eta \cdot (q_j-q_i)} \, ,
\label{eq:LTDexample}
\eeq
with $m_i$ the mass associated to the internal line with momenta $q_i=\ell+k_i$ ($k_i=p_1 + \ldots + p_i$) and $\eta$ an arbitrary future-like vector (i.e. $\eta^2\geq 0$). It is important to notice that the usual Feynman prescription in the l.h.s. of Eq. (\ref{eq:LTDexample}) is converted into a \emph{modified} prescription inside the dual propagators in the r.h.s., which depends on the momenta carried by the cut line and the propagating particle. Also, the loop measure is transformed into a phase-space measure by inserting the factor $\td{q_i}=2\pi \, \imath \, \theta(q_{i,0})\delta(q_i^2-m_i^2)$, that forces the momenta $q_i$ to represent a physical on-shell state with positive energy. 

Beyond one loop, the LTD theorem establishes that the number of cuts required to formulate the dual representation equals the number of loops involved \cite{Catani:2008xa,Bierenbaum:2010cy}. This is particularly important since N$^n$LO computations involve adding together all the possible $(n-l)$-loop amplitudes with $l$ radiated particles, where $l=0,\ldots,n$. So, after the iterative application of the LTD, any N$^n$LO calculation is reduced to a set of tree-level objects containing $n$ additional on-shell positive-energy momenta \cite{Sborlini:2016hat}. 

In the following sections, we will explain how to unveil the IR structure of the virtual amplitudes by using the LTD. Moreover, by exploiting this knowledge, we will define a set of kinematical transformations that allows to map the IR singular points of the dual and real contributions to the same points in the integration domain.

\begin{figure}[b]
\begin{center}
\includegraphics[width=0.35\textwidth]{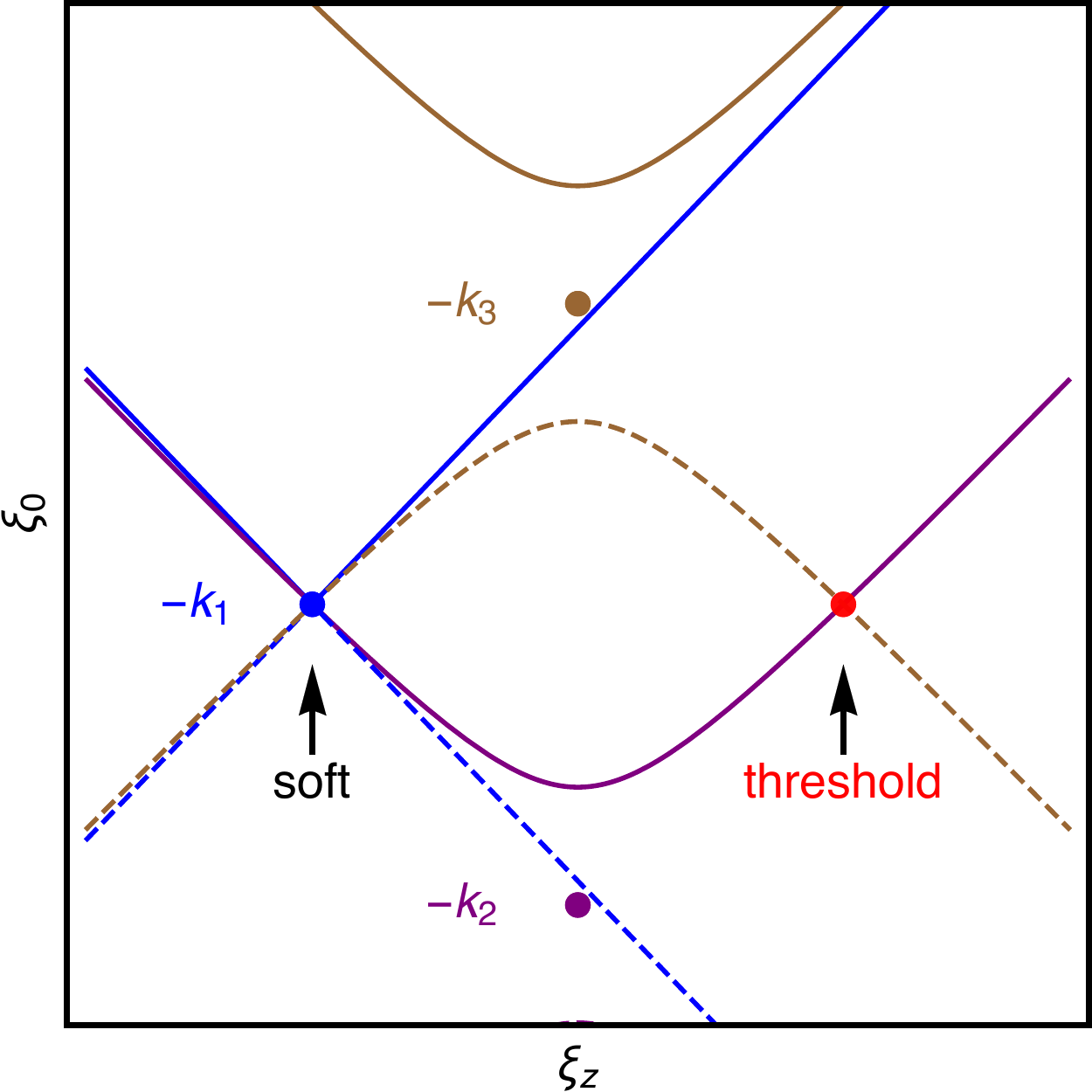} \ \ \ \ \ \ \ \ 
\includegraphics[width=0.35\textwidth]{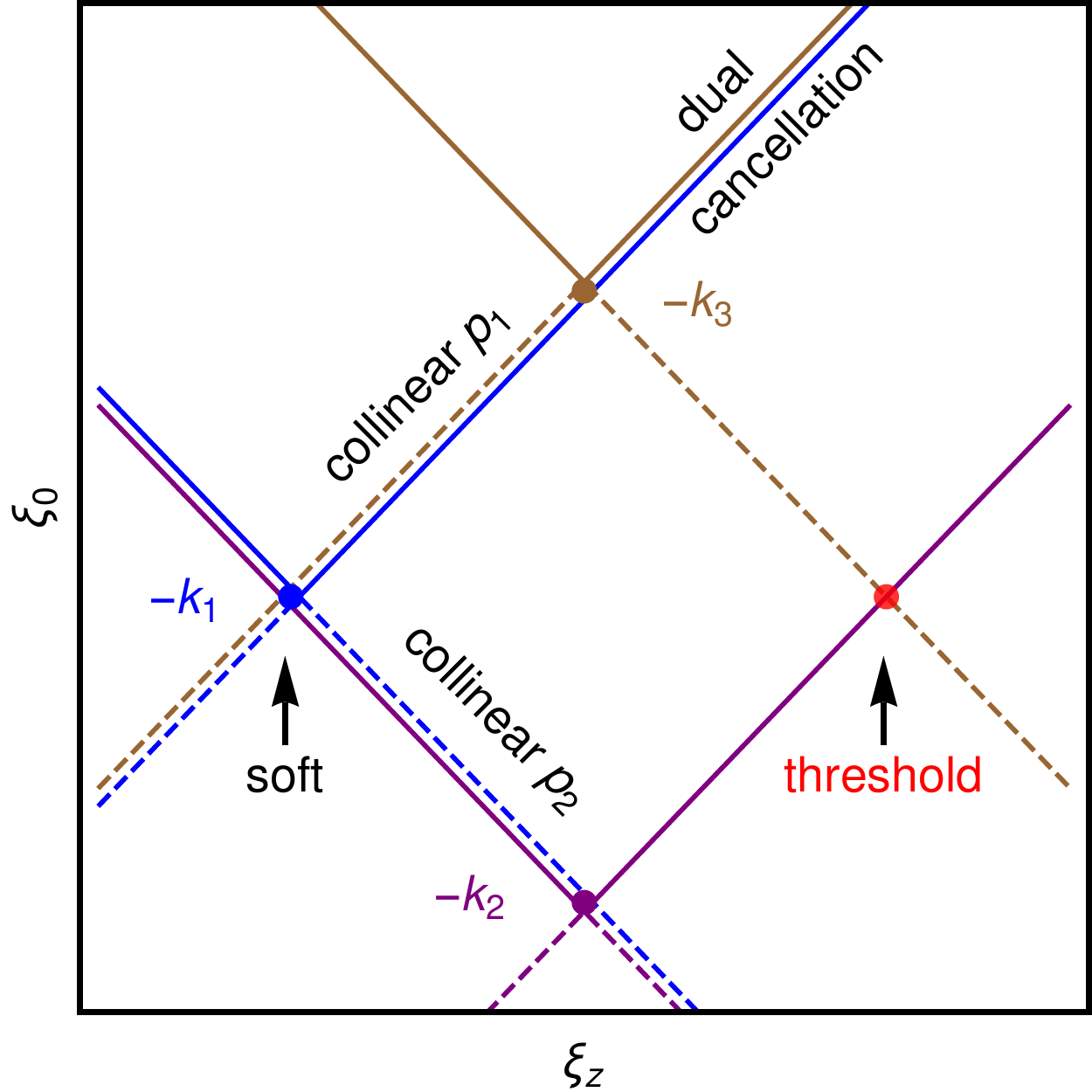}
\caption{\label{fig:IRSINGULARITIES}
Integration region in the $(\xi_0,\xi_z)$-plane for the dual contributions associated with a triangle with two massive lines (left) and with only massless particles (right). The solid (dashed) lines represent the forward (backward) regions, i.e. the positive (negative) energy solutions of the on-shell conditions. Intersection among these lines are related with multiple propagators becoming singular; in particular, forward-backward intersections lead to the IR poles of the Feynman integral.}
\end{center}
\end{figure}

\subsection{Location of IR singularities}
\label{ssec:IRsingularities}
After the application of the LTD, we get a set of \emph{dual} amplitudes which include a factor $\td{q_i}$ inside the integration measure. This means that we must restrict the integration to the solutions of the on-shell conditions, i.e. 
\beq
G_F^{-1}(q_i) = q_i^2-m_i^2 + \imath 0 = 0 \, , \quad q_{i,0}^{(+)} = \sqrt{\vec{q}_i^2+m_i^2-\imath 0} \, ,
\eeq
where we choose the positive energy solutions. In Fig. \ref{fig:IRSINGULARITIES}, we consider the integration domains associated to a Feynman integral with three propagators. In the left panel, the integral under consideration contains two internal lines with mass $M$ and a massless one, which translates into the presence of two hyperboloids for the massive particles and a light-cone associated to the massless line. When the limit $M \to 0$ is considered, the hyperboloids degenerate into light-cones, as shown in the right panel. In both cases, the solid (dashed) lines represent the forward (backward) regions, i.e. the positive (negative) energy solutions of the on-shell conditions. The crucial fact is that the intersection of the on-shell hyperboloids is related with the presence of IR and threshold singularities. Essentially, this is due to multiple particles satisfying the on-shell condition, and, thus, more than one propagator becoming singular in the integration domain. However, as explained in Refs. \cite{Buchta:2014dfa,Buchta:2015wna}, forward-forward intersections cancel among dual contributions but forward-backward intersections originate the physical IR poles. Punctual intersections are related with threshold singularities, that are integrable but might introduce some numerical instabilities.

We can appreciate in the right panel of Fig. \ref{fig:IRSINGULARITIES} that the region responsible of the IR singularities of the virtual contribution is contained in a compact domain. This is \emph{the} property that allows to infer how to combine the real and virtual contributions to achieve a local cancellation of IR singularities, as we carefully explained in Refs. \cite{Hernandez-Pinto:2015ysa,Sborlini:2016gbr,Sborlini:2016hat}. Since the IR structure of the virtual component matches the one in the real part, and the real-emission PS is finite, then regularizing the combined real-virtual contribution is equivalent to properly mapping the singular points in two compact spaces. 

\begin{figure}[ht]
\begin{center}
\includegraphics[width=0.55\textwidth]{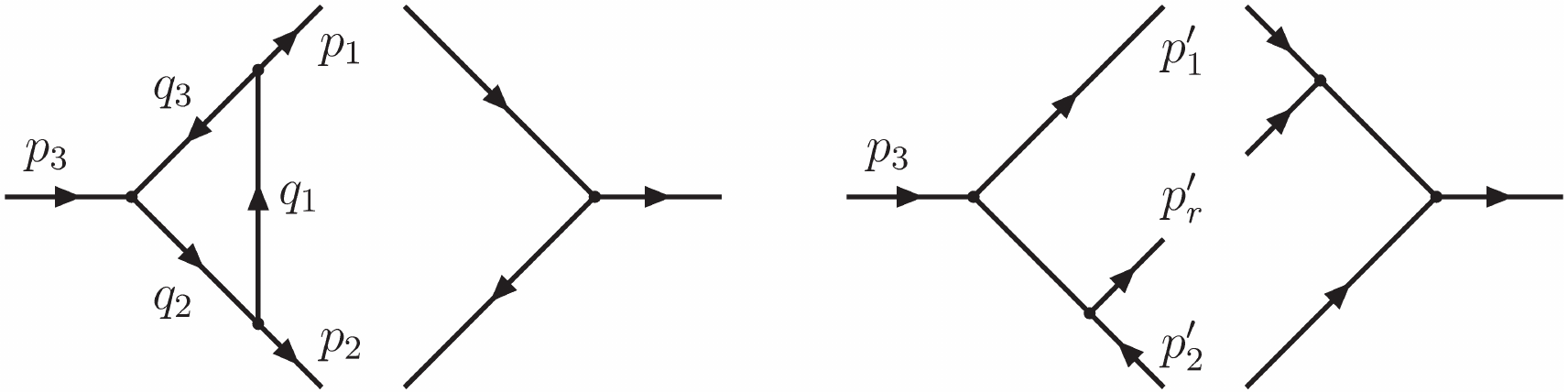}
\caption{\label{FACTORIZACION}
Topological correspondence among one-loop (left) and real-emission amplitudes (right). When we cut the line $q_2$ and consider the limit $q_1 \parallel p_1$, the virtual diagram factorizes in the same way that the real contribution does in the limit $p^{'}_r \parallel p^{'}_1$. This information is used to split the real-phase space and build the proper momentum mapping in each partition.}
\end{center}
\end{figure}

\subsection{Real-virtual mapping}
\label{ssec:RVmapping}
In order to motivate the construction of the real-virtual momentum mapping, let's consider that the Born process contains $m$ external momenta. Then, the NLO corrections are given by one-loop virtual amplitudes with $m$ on-shell momenta and a free loop-momentum $q$, and $m+1$ real-emission amplitudes with $m+1$ on-shell momenta. Once the LTD is applied to the virtual part, we obtain $m$ dual components, which are described in terms of the external $m$ momenta and a free \emph{on-shell} momentum with \emph{positive} energy. Thus, the number of kinematical variables in both contributions exactly matches and we can propose a transformation to connect them. We denote the Born level momenta as $\{p_i^{\mu}\}_{i=1\ldots m}$, $\vec{q_j}$ is the spatial part of the dual momentum and $\{p_i'^{\mu}\}_{i=1\ldots m+1}$ are the momenta entering in the real-emission process. 

In order to simplify the development of the mapping, we introduce a partition of the real phase-space to isolate the collinear configurations \cite{Frixione:1995ms}. Explicitly, we define ${\cal R}_i = \{{y'}_{i r} < {\rm min}\, {y'}_{jk} \}$, where ${y'}_{ij}=2\, p_i' \cdot p_j'/Q^2$, $r$ is the label associated to the radiated parton from parton $i$, and $Q$ is the typical hard scale of the scattering process. Of course, we have $m$ regions and the constraint $\sum_{i=1}^{m} {\cal R}_i =1$ to cover the whole phase-space. With this definition, the only allowed collinear/soft configurations in ${\cal R}_i$ are $i\parallel r$ or $p_r'^{\mu} \to 0$. 

The next step consists in connecting the dual contributions and the real-emission components inside an specific partition. Motivated by the factorization picture shown in Fig. \ref{FACTORIZACION}, we look for the diagrams which originate the same kind of topologies when their internal lines are put on-shell: the cut-line in the dual amplitude must be interpreted as the extra-radiated particle in the real contribution. This means that if we identify $q_i \leftrightarrow p_r'$, then the have to look for the real-emission diagrams that become singular in the limit $i \parallel r$, and restrict the integration to the region ${\cal R}_i$.  

After the previous explanations, let's present the explicit mapping. Using the language of the dipole formalism \cite{Catani:1996jh,Catani:1996vz}, if $i$ is the \emph{emitter} and $r$ is the \emph{radiated} particle, the transformation that generates the kinematics inside the partition ${\cal R}_i$ with the kinematics of the $i$-th dual component is given by 
\bea
&& p_r'^\mu = q_i^\mu~, \ \ \quad \quad \quad \quad \quad \quad \quad p_j'^\mu = (1-\alpha_i) \, p_j^\mu~, \nn \\
&& p_i'^\mu = p_i^\mu - q_i^\mu + \alpha_i \, p_j^\mu~, \qquad \alpha_i = \frac{(q_i-p_i)^2}{2 p_j\cdot(q_i-p_i)}~,
\label{generalmapping}
\eea
where the particle $j$ is the \emph{spectator}. Notice that, by construction, the generated momenta fulfill
$p_k'^2=0$ (in the massless case) and $\sum_l \, {p'}_l =0$. On the other hand, it is important to mention that the mapping does not change the initial-state momenta ($p_a$ and $p_b$) neither $p'_k$ with $k \ne i,j$, and that the global momentum conservation is respected.


\section{Application examples at NLO}
\label{sec:examplesNLO}
\begin{figure}[t]
\begin{center}
\includegraphics[width=0.43\textwidth]{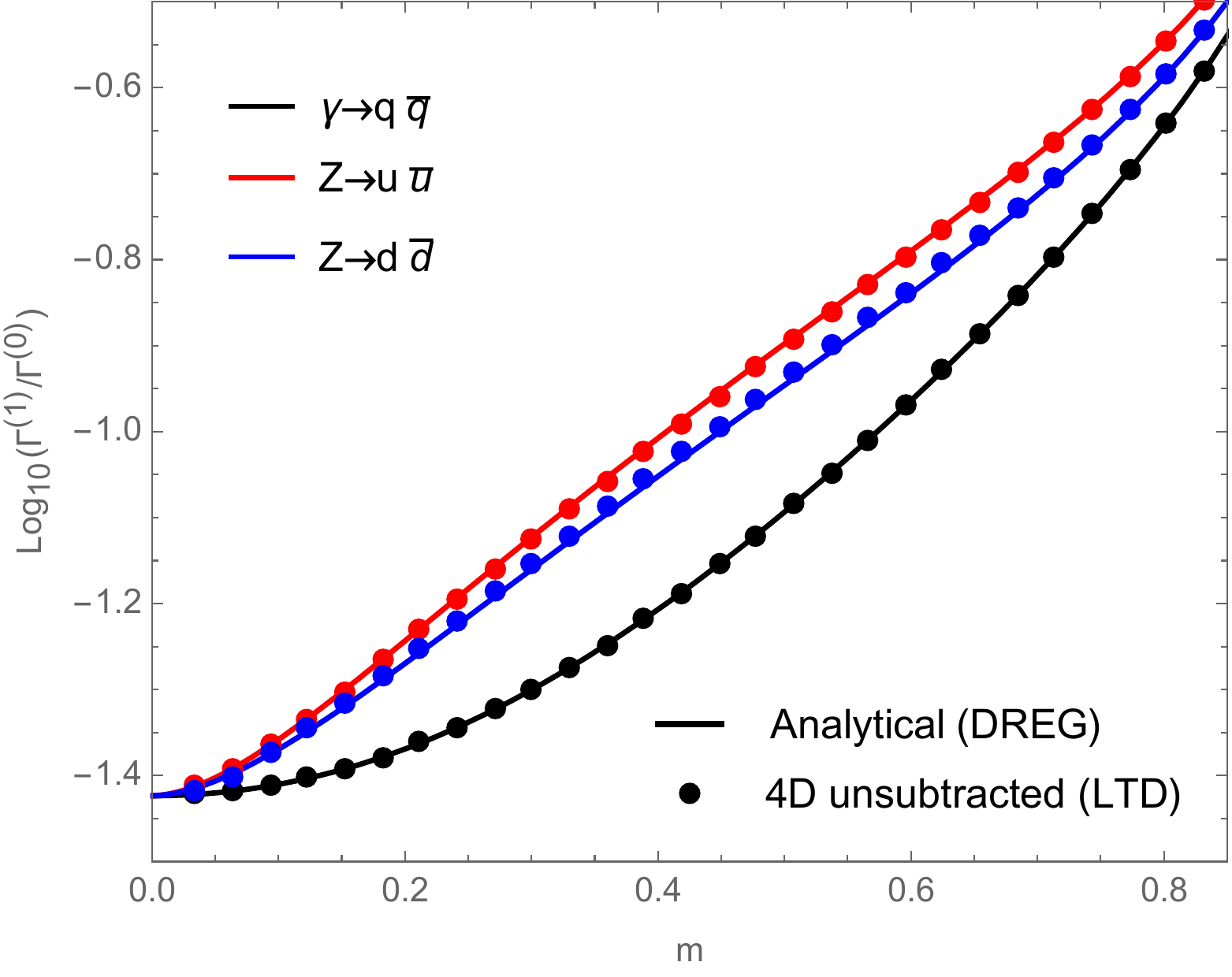} \ \ \ \ \ \ 
\includegraphics[width=0.42\textwidth]{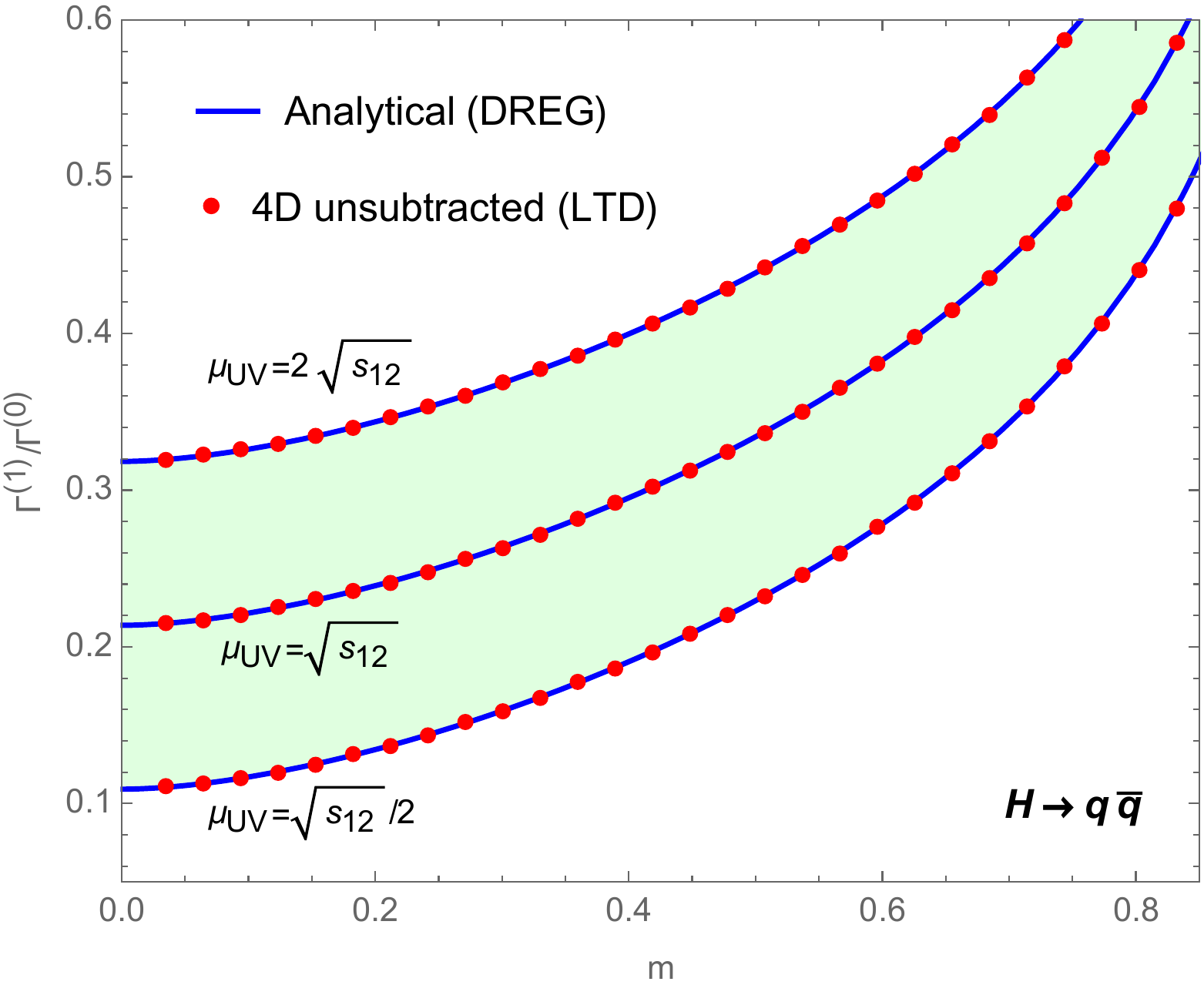}
\caption{\label{fig:NLOEXAMPLES}
NLO QCD corrections to the decay rates $Z,\gamma^* \to q \bar q$ (left) and $H \to q \bar q$ (right), as a function of the mass $m$ of the quarks. The solid lines represent the results obtained within the DREG approach, which are known in a closed analytical form. The colored dots were computed numerically thorough the application of the FDU technique. We can appreciate the agreement between these approaches, and in particular the smoothness of the massless limit. Moreover, the scale dependence for the Higgs decay is exactly reproduced, thanks to the introduction of local UV counter-terms.}
\end{center}
\end{figure}

We use the FDU approach to recompute the decay processes $H \to q \bar q$ and ${Z,\gamma^*} \to q \bar q$ at NLO in QCD with massive quarks \cite{Sborlini:2016hat}. In both cases, we apply the techniques mentioned in the previous sections to obtain a combined real-virtual integrand with a regular behavior, i.e. numerically integrable in four dimensions. The results are compared with the known expressions computed within the DREG framework. In Fig. \ref{fig:NLOEXAMPLES}, the solid lines denote the analytical results computed in DREG as a function of the quark mass $m$, whilst the colored dots are the values obtained through the FDU implementation. 

In first place, we appreciate that the agreement between both approaches is excellent. We notice that the massless transition is smooth in both cases, which is a non-trivial result. In fact, when dealing with the analytical expressions with $m>0$, we find some logarithmic-enhanced terms in the real and virtual contributions, separately. Within DREG, these logarithms transforms into $\epsilon$-poles when considering the limit $m \to 0$; thus, a naive implementation of the calculation could lead to huge numerical instabilities. It is a remarkable property of the FDU approach that this transition is completely stable numerically, as a consequence of the local regularization of the integrand and the smoothness of the real-virtual mapping in the massless limit.

Also, we should mention that these computations involve dealing with UV singularities. They are hidden inside the renormalization factors and the self-energy corrections \cite{Ramirez-Uribe:2017gbf}, as well as in the \emph{usual} virtual diagrams. In the processes under consideration here, there is a partial cancellation of UV divergences between the vertex corrections and the self-energy contributions (which also contains some IR-singular pieces). To achieve integrability in four-dimensions, we define local UV counter-terms and apply the LTD to obtain the corresponding dual expressions. The presence of higher-powers of the propagators inside these counter-terms is tackled with an extended version of the LTD, as discussed in Refs. \cite{Sborlini:2016gbr,Sborlini:2016hat,Bierenbaum:2012th}. These dual UV counter-terms are added to the remaining dualized virtual contributions, rendering the total sum finite in the high-energy limit. In fact, as we can appreciate from the right panel of Fig. \ref{fig:NLOEXAMPLES}, the renormalization scale dependence is successfully reproduced with our four-dimensional framework. 

\section{Conclusions and outlook}
\label{sec:conclusions}
In this article, we present some features of the four-dimensional unsubtraction (FDU) approach. This method is intended to implement higher-order computations of physical observables in any QFT, and is based on the loop-tree duality (LTD) theorem. The dual decomposition of finite Feynman integrals was proven to be very efficient for numerical calculations in four-dimensions \cite{Buchta:2015wna}, even when dealing with complicated tensorial structures. Thus, the following natural step was its extension to deal with any physical observable.

The complication of higher-order QFT computations is the presence of IR and UV singularities, which forces to introduce regularization methods and counter-terms. In the usual subtraction framework, these counter-terms cancel locally the IR divergences of the real components, but analytical manipulations are required for the virtual part (as well as the renormalization). Recently, there were many developments to by-pass DREG with alternative regularization techniques, leading to integrable expressions in four-dimensions \cite{Gnendiger:2017pys}. It is worth appreciating that regularization might be used even when the final result is finite, since intermediate steps could contain ill-defined expressions \cite{Driencourt-Mangin:2017gop,Sborlini:2017mhc}.

In conclusion, the FDU approach allows to combine all the ingredients required to perform higher-order computations into a single numerically-integrable expression. Moreover, since this approach is completely process-independent, it could be used to develop fully-automated numerical implementations for any physical observable, without dealing with complicated analytical formulae in intermediate steps.

\section*{Acknowledgments}
This work is partially supported by the Spanish Government, by EU ERDF funds (grants FPA2014-53631-C2-1-P and SEV-2014-0398), by Generalitat Valenciana (GRISOLIA/2015/035) and Fondazione Cariplo under the grant number 2015-0761.


\end{document}